
\magnification=\magstep1
\baselineskip=14 pt
\hsize=5 in
\vsize=7.3 in
\pageno=1
\centerline{\bf PARTICLE CREATION AMPLIFICATION IN CURVED SPACE}
\centerline{\bf DUE TO THERMAL EFFECTS}
\vskip 1,8cm
\centerline{\bf Carlos E. Laciana}
\centerline {\sl Instituto de Astronom\'{\i}a y F\'{\i}sica del Espacio}
\centerline{\sl Casilla de Correo 67 - Sucursal 28, 1428 Buenos Aires,
Argentina}
\centerline{\sl E-mail: laciana@iafe.uba.ar}

 A physical system composed by a scalar field minimally coupled to
gravity and  a thermal reservoir, as in thermo field dynamics, all of then in
curved space time, is considered. When the formalism of thermo field
dynamics is generalized to the above mentioned case, an amplification in the
number of created particles is predicted.
\vskip 2,5cm
1. {\bf Introduction}

 As is shown in ref.[1] Thermo Field Dynamics (TFD) gives us, for thermal
equilibrium at a given temperature, an elegant method to obtain mean values
of physical magnitudes resulting at this level equivalent to the Schwinger-
Keldysh closed time path formalism (SKF). However as is shown in ref.[2]
TFD has advantages on the SKF formalism when the generalization to
nonequilibrium situations is performed, giving different results.

 The main idea in TFD is the duplication of the degrees of freedom,
introducing  the  tilde  operators    ${{\tilde  a}_{\bf  k}}$  and  ${{{\tilde
a}^\dagger}}_{\bf k}$ which operate on the quantum states of the reservoir
[3]. In our generalization the system and the reservoir are in a curved
background. The operators ${a_{\bf k}}$, ${{a^{\dagger}}_{\bf k}}$,
${\tilde a}_{\bf k}$ and ${{{\tilde a}^{\dagger}}_{\bf k}}$ are all related
to the thermal operators ${a_{\bf k}}(\beta)$,
${{a^{\dagger}}_{\bf k}}(\beta)$,
${{\tilde a}_{\bf k}}(\beta)$ and ${{{\tilde a}^{\dagger}}_{\bf k}}(\beta)$
by means of a Bogoliubov  transformation.    Moreover due to the curvature of
the space-time we have a set  of  vacua  related  to  the  folliation.    The
operators related with those vacua are indicated by ${a_{\bf k}}(t)$,
${{a^{\dagger}}_{\bf k}}(t)$,
${{\tilde  a}_{\bf  k}}(t)$ and ${{{\tilde a}^{\dagger}}_{\bf k}}(t)$.    The
parameter  ``$t$"  labels  the  folliation.    We  have   also  a  Bogoliubov
transformation  that  relate  those  operators  with the primitive ones  (see
ref.[4]). The particle creation will be due to two effects, one of them the
interaction  with the geometry and the other one related to  the  interaction
with the thermal bath.  The first one is a dynamical  effect  and the second
one is a spontaneous phenomenon.

\bigskip
2. {\bf Thermal creation of particles}

 Similarly as in  ref.[5], we can relate operators
at zero temperature with
the ones at temperature $T=T(\beta)$ by means of the following time
dependent transformation matrix:
\smallskip
$${a(\beta)\choose {{\tilde a}^{\dagger\!\dagger}}(\beta)}=
{(1+{n_{0 \beta}})^{1/2}}\pmatrix {1&{-F}\cr {-fF^{-1}}&1}
{a\choose {{\tilde a}^{\dagger}}}\eqno (2.1)$$
\smallskip
where $a_{\bf k}$ (${{a^{\dagger}}_{\bf k}}$) are quasiparticle annihilation
(creation) operators for bosonic fields
(in equation (2.1) we omitted the subindex {\bf k}
asociated with the mode of the quasiparticle),
defined by
\smallskip
$$a_{\bf k}|0>=0.$$
$${{a^{\dagger}}_{\bf k}}|0>=|1_{\bf k}>, etc$$
$$[{a_{\bf k}},{{a^{\dagger}}_{\bf k^{\prime}}}]=
{{\delta}_{{\bf k},{\bf k^{\prime}}}}$$
\smallskip
The tilde operators ${\tilde a}_{\bf k}$, ${{\tilde a}^{\dagger}}_{\bf k}$
represent the quantum effect of the reservoir. We have also
$${{\tilde a}_{\bf k}}|\tilde 0>=0. $$
$${{{\tilde a}^{\dagger}}_{\bf k}}|\tilde 0>=|{\tilde 1}_{\bf k}>,\  etc$$
$$[{{\tilde a}_{\bf k}},{{{\tilde a}^{\dagger}}_{\bf k^{\prime}}}]=
{{\delta}_{{\bf k},{\bf k^{\prime}}}}$$

 We will use  $\{|n,\tilde n>\}=
\{|n>\}\otimes\{|\tilde n>\}$ as space of states.
 In the following we will call $|{\bf 0}>:=|0,\tilde 0>$.

 The operators $a(\beta (t))$, ${a^{\dagger}}(\beta (t))$,
${\tilde a}(\beta (t))$ and ${\tilde a}(\beta (t))$ operate on the thermal
vacuum:
$${a_{\bf k}}(\beta)|{\bf 0},\beta>=0.$$
$${{a^{\dagger\!\dagger}}_{\bf k}}(\beta)|{\bf 0},\beta>
=|{1_{\bf k}},\beta>, \ etc$$

and satisfy

$$[{a_{\bf k}}(\beta), {a^{\dagger\!\dagger}}_{{\bf k}^{\prime}}(\beta)]=
{\delta }_{{\bf k}{{\bf k}^{\prime}}}$$

and similarly the operators ${{\tilde a}_{\bf k}}(\beta)$ and
${{{\tilde a}^{\dagger\!\dagger}}_{\bf k}}(\beta)$.
The symbol $\dagger\!\dagger$ is
used because ${(a^{\dagger\!\dagger})^{\dagger}}\ne a$ i.e. in general the
transformation is not unitary (only when $F=f^{1/2}$).

 Now we will define the function ${n_{0\beta}}(t)$ as the mean value of
created particles due to the interaction with the reservoir, in the form
$$n_{0\beta}=<{\bf 0}|{a^{\dagger\!\dagger}}(\beta) a(\beta)|{\bf 0}>
\eqno (2.2)$$
\smallskip
  As we can see using eq. (2.1)
$$n_{0 \beta}={f\over {1-f}}$$
\vskip 0,2cm
$f$ is in principle an arbitrary time dependent function, in particular
when $f=\exp {-{\beta}{\epsilon}}$ we have planckian spectrum.

 The time dependent function $F$ is also arbitrary. One choise used in the
literature [6] is known as thermal state condition, which corresponds to
$F=f^{\alpha}$ with $0<{\alpha}<1$. In particular when $\alpha = 1/2$,
 ${{(a^{\dagger\!\dagger})}^{\dagger}}=a$
(then ${\dagger\!\dagger}=
\dagger$) and the transformation is unitary.

 It is easier to prove also, using the inverse transformation:
$$<{\bf 0}|{a^{\dagger\!\dagger}}(\beta)a(\beta)|{\bf 0}>=
<{\bf 0},\beta|{a^{\dagger\!\dagger}}a|{\bf 0},\beta>$$
\bigskip
3. {\bf Particle creation in curved space}

 Following Parker (ref. [4]) we can relate the annihilation-creation
operators at different times, by means of the following Bogoliubov
transformation:
$${a_{\bf  k}}(t)={e^{i{{\gamma}_{\alpha}}({\bf    k}, t)}}cosh
{{\theta}({\bf
k},t)}{a_{\bf  k}}+{e^{i{{\gamma}_{\beta}}({\bf k},  t)}}sinh  {{\theta}({\bf
k},t)}{a^{\dagger}}_{-{\bf k}}$$

$${{a^{\dagger}}_{\bf  k}}(t)={e^{-i{{\gamma}_{\beta}}({\bf    k}, t)}}
sinh {{\theta}({\bf
k},t)}{a_{\bf  k}}+{e^{-i{{\gamma}_{\alpha}}({\bf k}, t)}}cosh {{\theta}({\bf
k},t)}{a^{\dagger}}_{-{\bf k}}\eqno (3.1)$$
\vskip 0,2cm

 Where  ${{\gamma}_{\alpha}}$,    ${{\gamma}_{\beta}}$    and   $\theta$  are
determined by the particle model used and by the field equation. When an
isotropic Robertson-Walker metric is considered the functions introduced
in eq. (3.1) satisfy the eqs. (see eqs (27) and (39) of ref.[4]):

$${{\dot \gamma}_{\beta}}{\tanh {\theta}}{\cos {\gamma}}+{{\dot {\theta}}
\sin {\gamma}}-{{{\dot \gamma}_{\alpha}}{\cos {\mu}}}+{\dot \theta}
{\tanh \theta}{\sin \mu}=0$$

$${-{{\dot \gamma}_{\beta}}}{\tanh {\theta}}{\sin {\gamma}}+{{\dot {\theta}}
\cos {\gamma}}-{{{\dot \gamma}_{\alpha}}{\sin {\mu}}}+{\dot \theta}
{\tanh \theta}{\cos \mu}=0$$

with $\mu :=2{{\int }_{t_{0}}}^{t}W({\bf k}, {t^{\prime}})d{t^{\prime}}$ and
$\gamma :={{\gamma}_{\alpha}}+{{\gamma}_{\beta}}$.  Therefore in eqs (3.1) we
 have six    unknowed    functions: ${{\dot    \gamma}_{\beta}}$,    ${{\dot
\gamma}_{\alpha}}$,
$\gamma$, $\theta$, $\dot \theta$ and $\mu$.  Two initial conditions are
added  to the two equations (3.1).
In reference [4] those conditions
are the following
$$\theta ({\bf k}, 0)=0$$
$${{\gamma}_{\alpha}}({\bf k}, 0)=0\eqno (3.2)$$
\vskip 0,2cm
which give us  as initial conditions a particular functional
form (similar to WKB) for the field modes.
  To completely solve the
problem two additional conditions must be introduced which determine
definitively, at all instant, the particle model.

 The
annihilation-creation operators of eq. (3.1) satisfy the bosonic commutation
relations, i.e. :
$$[{a_{\bf k}}(t), {a_{{\bf k}^{\prime}}}(t)]=0, [{{a^{\dagger}}_{\bf k}}(t),
{{a^{\dagger}}_{{\bf k}^{\prime}}}(t)]=0,$$
and
$$[{a_{\bf k}}(t), {{a^{\dagger}}_{{\bf k}^{\prime}}}(t)]={{\delta}_{{\bf k},
{{\bf k}^{\prime}}}} \  \ \forall \ \ t$$
\vskip 0,2cm
as we can see from eqs. (3.1) and the initial condition (3.2)
$${a_{\bf k}}(t=0)= a_{\bf k}$$
$${{a^{\dagger}}_{\bf k}}(t=0)={a^{\dagger}}_{\bf k}$$
$${a_{\bf k}}|0>=0,\ \ \ {{a^{\dagger}}_{\bf k}}|0>= |1_{\bf k}>$$
$${a_{\bf k}}(t)|0,t>=0,  \  \  \  {{a^{\dagger}}_{\bf  k}}(t)|0,t>=  |1_{\bf
k}>.$$

 We also define the mean value of created particles with ${\bf k}$ mode by
${{n^{\bf k}}_{0t}}=<0|{{a^{\dagger}}_{\bf k}}(t){a_{\bf k}}(t)|0>$. In the
following, the ${\bf k}$ index is supressed.

\bigskip

4. {\bf Introduction of tilde operators in curved space-time}

 Next we will assume that the system and the reservoir are both
in a curved background.

 In order  to  use  the transformations given by eqs.  (2.1) and (3.1) on the
same operators, we will introduce the quadrivectorial operators $A_{\bf k}$,
${A_{\bf k}}(\beta)$ and ${A_{\bf k}}(t)$, defined by
(using $l$ as generic variable)

$${{\bf A}_{\bf k}}(l):=\pmatrix{{a_{\bf k}}(l)\cr
                                    {{a^{\dagger}}_{\bf k}}(l)\cr
                                      {{\tilde a}_{\bf k}}(l)\cr
      {{{\tilde a}^{\dagger}}_{\bf k}}(l)\cr}\eqno (4.1)$$

 Let us also introduce the  $4\times4$ matrices ${\bf \Omega}$ and
${\bf \Upsilon}$ so
that the two transformations can be written as
$${{\bf A}_{\bf k}}(t)={\bf \Omega}({\bf k},t){{\bf A}_{\bf k}}\eqno (4.2a)$$
$${{\bf A}_{\bf k}}(\beta)={\bf \Upsilon}({\bf k},\beta){{\bf A}_{\bf k}}\eqno
(4.2b)$$
\vskip 0,2cm
where
\vskip 0,1cm
$${\bf \Omega}=\pmatrix{ {\bf M}&{\bf 0}\cr {\bf 0}&{\bf M}\cr}$$
$${\bf M}=\pmatrix{  {e^{{i{\gamma}_{\alpha}}({\bf  k},t)}{\cosh{\theta}({\bf
k},t)}},
&{{e^{{i{\gamma}_{\beta}}({\bf  k},t)}{\sinh{\theta}({\bf
k},t)}}{\bf P}
}\cr {{e^{{-i{\gamma}_{\beta}}({\bf  k},t)}{\sinh{\theta}({\bf
k},t)}}{\bf P},
}&{{e^{{-i{\gamma}_{\alpha}}({\bf  k},t)}{\cosh{\theta}({\bf
k},t)}}
}\cr}$$
$${\bf \Upsilon}={\pmatrix{ {\bf I}&{\bf L}\cr {\bf L}&{\bf I}\cr}}(1+{n_{0
\beta}})^{1/2}$$
$${\bf L}=\pmatrix{ {0}&{{-F}}\cr {{-f{F^{-1}}}}&{0}\cr}$$
Moreover ${\bf I}$ is the $2\times2$ identity matrix and ${\bf 0}$
the zero $2\times2$
matrix. ${\bf P}$ is a parity operator, such that
$${\bf P}{a_{\bf k}}= a_{-{\bf k}}$$

 The total number of created particles comes from two sources,
a geometric one due to
the curvature of the space-time and the other one due to the thermal effects.
Therefore we can define $n_{0t\beta}$ as the mean value of the total
number of created particles, by
$${n_{0t\beta}}=<{\bf 0}|{{a^{\dagger\!\dagger}
}(t,{{\beta}(t)})}a(t,{\beta}(t))|{\bf 0}>\eqno(4.3)$$
\vskip 0,1cm

 Then we need the transformation
$${{\bf A}(t,{\beta}(t))}={{\bf \Upsilon}(t)}{\bf \Omega}(t){\bf A}:=
{{\bf \Lambda}(t)}{\bf A}$$

 therefore

$${\bf \Lambda}={(1+{n_{0\beta}})^{1/2}}{(1+{n_{0t}})^{1/2}}\pmatrix{{\bf R}&
{\bf T}\cr
{\bf T}&{\bf R}}$$
\vskip 0,1cm
with
$${\bf R}=\pmatrix{e^{i{{\gamma}_{\alpha}}}&{e^{i{{\gamma}_{\beta}}}}
{\tanh \theta}{\bf P}\cr {e^{-i{{\gamma}_{\beta}}}}{\tanh \theta}{\bf P}&
e^{-i{{\gamma}_{\alpha}}}},$$
$${\bf T}=\pmatrix{-F{e^{-i{{\gamma}_{\beta}}}}{\tanh \theta}{\bf P}&-F
e^{-i{{\gamma}_{\alpha}}}\cr -f{F^{-1}}e^{i{\gamma}_{\alpha}}&-f{F^{-1}}
{e^{i{{\gamma}_{\beta}}}}{\tanh \theta}{\bf P}}$$

 When the calculation in eq. (4.3) is performed the following result is
obtained
$${n_{0t\beta}}={n_{0t}}+2{n_{0\beta}}{n_{0t}}\eqno (4.4)$$
\vskip 0,1cm
This result is independent of the particle model. From eq.(4.4) we see that
if at  the  begining we had a curved geometry at zero temperature there
would be a
number $n_{0t}$ of created particles at time ``$t$".  When all the system
is in interaction with a reservoir at temperature $T$ the number of particles
increases in agreement with eq. (4.4).
\bigskip
5. {\bf Concluding remarks.}

 As was shown in ref.[4] the transformation given by eq.(3.1) is a consistent
way to obtain Heinsenberg annihilation-creation operators for a
Robertson-Walker metric.  For other metrics the Bogoliubov transformation may
be different. However the result obtained in eq.(4.4) is independent of the
particle model used.

 The increment in the number of created particles due to the
temperature is equal to
the one produced in curved space at zero temperature, when there is
a particle distribution  in  the  initial  state,  as  is  proved in
ref.[4].  In other words  this  fact  is  related  with the analogy, which is
shown in ref.[7], between the role of the tilde operator ${\tilde a}_{\bf k}$
and the operator $a_{-{\bf k}}$ in curved space.

 In our generalization of TFD to a curved space we  said nothing about
what the tilde modes are.    We  think  that the role of tilde modes is
played  by  the  quantum  perturbations  of the metric,  this  possibility  is
presently being studied by us.
 As a first approach  we  found  that  the conformal
fluctuation of the metric can satisfy all the hypothesis in order to act as a
natural tilde field (see ref.[8]).
\vskip 1,8cm
\noindent
{\bf ACKNOWLEDGEMENTS}
\medskip

 This work was supported by the European Community DG XII and by the
Departamento de F\'{\i}sica de la Fac. de Ciencias Exactas y Nat. de
Buenos Aires.
\vfill\eject
\vskip 1cm
{\bf REFERENCES}
\medskip
\item{$[1]$} Takahashi Y. and Umezawa H.; Collective Phenomena {\bf 2},
55-80, (1975).
\item{$[2]$} Chu H. and Umezawa H.; Int. J. of Modern Phys. A, {\bf 9},
$N_{o} 14$ (1994) 2363-2409.
\item{$[3]$} Umezawa H., Matsumoto H., and Tachiki M. (1982); ``Thermo
Field Dynamics and Condensed States" (North-Holland, Amsterdam).
\item{$[4]$} Parker L.; Phys. Rev. {\bf 183}, $N_{o} 5$, 1057-1068, (1969).
\item{$[5]$} Hardman I., Umezawa H., and Yamanaka Y.; J. Math. Phys.
{\bf 28} (12), (1987) 2925-2938.
\item{$[6]$} Arimitsu T., Guida M., and Umezawa H.; Physica {\bf 148 A}
(1988), 1-26.
\item{$[7]$} Laciana  C.E.;    Gen.    Rel.   and Grav.  {\bf 26}, $N_{o} 4$,
363-379, (1994).
\item{$[8]$} Laciana C.E.;    ``Can  gravity  be  a thermal reservoir",
Physica {\bf A} (in press).

\bye